# Effect of Rhombohedral to Orthorhombic Transition on Magnetic and Dielectric Properties of La and Ti co-substituted $BiFeO_3$

Pawan Kumar, Chandrakanta Panda and Manoranjan Kar*

Department of Physics, Indian Institute of Technology Patna, Patna-800013, India.

* Corresponding author Email: mano@iitp.ac.in, Ph: +916122552013, Fax: +916122277383

**Abstract:**

Polycrystalline $Bi_{1-x}La_xFe_{1-x}Ti_xO_3$ (x = 0.000 – 0.250) ceramics were synthesized by the tartaric acid modified sol-gel technique. It was observed that the co-substitution of La & Ti at Bi & Fe sites in $BiFeO_3$ suppress the impurity phase formation which is a common problem in bismuth ferrite. The quantitative crystallographic phase analysis was performed with the help of FULLPROF program which suggests the existence of compositional driven crystal structure transition from rhombohederal (space group *R3c*) to the orthorhombic (space group *Pbnm*). The changes in the phonon frequencies as well as line widths of $A_1$ mode in Raman spectra reveal the lattice distortion which tends to modify the crystal structure. The structural transition breaks the spin cycloid structure in co-substituted $BiFeO_3$ nanoparticles which leads to canting of the antiferromagnetic spin structure. Hence, the remnant magnetization increases up to 10 % of co-substitution and becomes 22 times that of $BiFeO_3$. However, it decreases for higher co-substitution percentage due to significant contribution from the collinear antiferromagnetic ordering in the orthorhombic crystal symmetry. The co-substitution significantly enhanced the dielectric constant (maximum in x = 0.100) as well as frequency independent region for dielectric constant and dielectric loss.

## 1. Introduction

Multiferroic materials have attracted a lot of attention due to their potential applications in memory devices, spintronic devices, magnetically modulated transducers, ultrafast optoelectronic devices and sensors etc. [1-4]. $BiFeO_3$ (BFO) is the most studied one because it exhibits ferroelectric order as well as G-type antiferromagnetic orderings above room temperature ($T_c$ = 810-830 $^oC$ and $T_N$ = 350-370 $^oC$) [5]. Its crystal structure is described by rhombohederally distorted perovskite structure with R*3c* space group. The ferroelectricity in this compound arises due to the off-centre structural distortions of cations induced by the $6S^2$ lone pair electrons of $Bi^{3+}$ whereas the G-type antiferromagnetic ordering arises due to indirect exchange interaction between $Fe^{3+}$ ions through $O^{2-}$ ions. The R3c symmetry allows the existence of weak ferromagnetic moment due to Dzyaloshinky-Moriya interaction but the cycloid spin structure with the periodicity of ~62 nm prevents net magnetization which leads to net zero magnetization [6-9]. Pure BFO undergoes antiferromagnet to weak ferromagnet phase transformation at H ~ 20 T [10]. Lanthanide substitution in BFO is known to reduce the threshold field of the magnetic phase transformation which is associated with the substitution-driven increase of the magnetic anisotropy [10, 11]. The substitution can affect the anisotropy directly by modifying structural factor (i.e. by changing the crystal field of ligands through the substitution with ions having a different ionic radius), or indirectly, via magnetoelectric coupling [12]. It has been reported that the high ferroelectric and ferromagnetic polarization or large magnetoelectric coupling constant at room temperature through A-site and/or B-site substitution in BFO [13-18]. Direct evidence of cycloid suppression in Lanthanum ($La^{3+}$) substituted ceramics has been given through nuclear magnetic resonance (NMR) measurements which has been correlated with the coexistence of rhombohedral to orthorhombic crystal structures [11, 13, 17]. This structural phase transition significantly enhances the magnetization as well as magnetoelectric interaction.

BFO has limited technological applications due to high leakage current at room temperature, which originate from high defect density, oxygen off-stoichiometry and presence of impurity phases ($Bi_2Fe_4O_9$ and $Bi_{25}FeO_{40}$ etc.). The creation of oxygen vacancies in BFO is due to the interconversion from $Fe^{3+}$ to $Fe^{2+}$. The literature survey suggests that the donor substituents reduce the concentration of oxygen vacancies by compensating for $Fe^{2+}$, thereby reducing the number of charge carriers. Several donor substituents have been tried to reduce conductivity in bismuth ferrite ceramics and thin films [18, 19]. Instead, $Ti^{4+}$ was chosen as the substituent because this is known to readily react with $Bi_2O_3$ [20] and is a highly polarizable ion which might couple with Bi displacements. It also breaks cycloid spin structure which leads to enhancement in the resultant magnetic moment. The substitution at the $Fe^{3+}$ sites increases the Fe-O-Fe angle and improves the magnetic properties because the increase in bond angle can increase the super-exchange interaction between the two Fe ions [21].

It is difficult to prepare single phase BFO because of its narrow temperature range of phase stabilization. Several attempts have been made to prepare phase pure by chemical route and the solid state route followed by leaching with nitric acid [22, 23]. The nitric acid leaching is used to eliminate above impurity phases which leads to the formation of coarser powders and its poor reproducibility. Moreover, Pradhan et al. prepared pure BFO phase ceramics by a rapid liquid-phase sintering technique. The crystallization temperature of BFO in these methods was above the ferroelectric Curie temperature $T_c$, which suggests that there will be volatilization of bismuth which deteriorates its dielectric properties [24]. Hence, we have adopted the chemical route of synthesis for uniform particle size and better reproducibility. Cheng et al. [25] have obtained $Bi_{1-x}La_xFeO_3$ ceramics with x = 0, 0.1, 0.2, and 0.3 by the solid-state reaction, starting from metal oxides. They observed that $Bi_{1-x}La_xFeO_3$ maintains the rhombohedral structure of BFO up to 10% of La substitution. However, the structures change to the orthorhombic for 20% substitution. They observed that La substitution significantly reduces electric leakage and enhances the ferromagnetic moment due to the broken cycloid spin structure caused by the changes in the crystalline structure. Also, M. Kumar et al. have reported that there is increase in resistivity and magnetoelectric coupling by substitution of Ti in B-site of BFO [26]. Chia-Ching Lee et. al and Jiagang Wu et. al. have observed that the La and /or Ti substitution has significantly reduced leakage current even in high electric field of the order of 1800 kV/cm. They also observed almost fatigue-free behavior was observed up to $10^9$ switching cycles [27, 28]. However, there is no report on influence of crystal structure transition on magnetic, optical and dielectric properties on La and Ti co-substituted BFO ceramics. As BFO is potential candidate for multiferroic device applications, it is needed to explore the correlation of different crystal symmetries with the above physical properties.

## 2. Methods

$Bi_{1-x}La_xFe_{1-x}Ti_xO_3$ with x = 0.000, 0.015, 0.025, 0.050, 0.075, 0.100, 0.150, 0.200 and 0.250 (named as BFO, BLFT-015, BLFT-025, BLFT-05, BLFT-075, BLFT-10, BLFT-15, BLFT-20 and BLFT-25 respectively) were synthesized by the tartaric acid modified sol-gel technique [29]. Here, bismuth nitrate, iron nitrate, lanthanum acetate, titanium oxide and tartaric acid (purity $\geq$ 99.0%) were the starting materials and carefully weighed in stoichiometric proportion. The resulting material was thoroughly grinded and annealed at 700 $^oC$ for 3 hours.

The crystallographic phases analysis of all the ceramics were carried out by the powder X-ray diffraction (XRD) study using 18 kW Cu-rotating anode based Rigaku TTRX III diffractometer, Japan) with CuKα radiation ($\lambda$ = 1.5418 Å) operating in the Bragg-Brentano geometry in a 2θ range of $10^o$-$120^o$ at a scan step of $0.01^o$ (counting time for each step was 2.0 Sec). Room temperature Raman spectra were measured in the backscattering geometry using confocal micro-Raman spectrometer (Seki Technotron Corp., Japan) with the 514.5 nm laser line as excitation source by STR 750 RAMAN Spectrograph using a 100 X microscope. The microstructural properties of the ceramics were investigated by using Field Emission Scanning Electron Microscopy (FE-SEM), Hitachi S-4800, JAPAN, operating at an accelerating voltage of 10 kV and equipped with energy-dispersive X-ray spectroscopic capability. The particle size and selected area electron diffraction (SAED) patterns were observed with the help of transmission electron microscopy (TEM, JEOL) at an accelerating voltage of 200 kV. Fourier Transform Infrared Spectra (FT-IR) were recorded at room temperature using the Perkin Elmer (model 400) in the range from 400 to 1200 $cm^{-1}$. The band gaps for all ceramics have been investigated by diffuse reflectance spectroscopy using LAMBDA 35 UV-visible spectrophotometer in the range from 200 to 1100 nm. Room temperature magnetization – magnetic field (M-H) measurement for all the ceramics were carried out by magnetic properties measurement system (MPMS), Quantum Design Inc., USA in the applied magnetic field of maximum ± 90 kOe. The parallel plane surfaces of pellets were coated with the silver paste for dielectric measurement. The dielectric constant and loss measurement were performed using the impedance analyzer (N4L PSM 1735) in the frequency range of 300 Hz – 8 MHz at the room temperature (30 $^oC$).

The XRD patterns were analyzed by employing the Rietveld refinement technique with the help of FULLPROF program [30]. The patterns for all the ceramics could be refined using the *R3c* as well as *Pbnm* space groups. Different structural parameters, such as, Zero correction, scale factor, half width parameters, lattice parameters, atomic fractional position coordinates, thermal parameters were varied during refinement. Background and peak shape was refined by the sixth order polynomial and pseudo-voigt function respectively.

**3. Results and discussion**

XRD patterns for $Bi_{1-x}La_xFe_{1-x}Ti_xO_3$ ceramics with x = 0.000 – 0.250 annealed at 700 $^oC$ are shown in the Fig. 1. Small impurity phases such as $Bi_2Fe_4O_9$ and $Bi_{25}FeO_{40}$ have been observed from XRD pattern for the samples x < 0.05. This is due to both anion and cation vacancies which were observed for BFO ceramics [31]. These impurities phases are suppressed with the increase in co-substitution of La & Ti. The (104) and (110) reflections are clearly separated in the BLFT-05 ceramics as shown in the Fig. 2. On increasing the La and Ti content, all the doublets appears merging to give a single peak, which is clearly visible ) for more than 5 % of co-substitution. These results correspond to the increase in lattice distortion for co-substituted ceramics with the increase in the substitution concentration which leads to rhombohedral to orthorhombic crystal phase transition. The lattice constants decreases with the increase in the substituent concentration due to smaller radius of substituents than that of host cations, which is evident from the shift of the XRD peak towards higher 2θ value as shown in the Fig. 2. This shift indicates that the substituents get substituted in the BFO lattice. The particle size has been calculated using Scherrer's formula [32] which is defined as,

$$D = \frac{k\lambda}{\beta Cos\theta} \quad \text{------------- (1)}$$

Where constant $k$ depends upon the shape of the crystallite size (= 0.89, assuming the spherical particles), $\beta$ = Full width at Half Maximum (FWHM) of Intensity (a.u.) vs. 2θ profile, $\lambda$ is the wavelength of the $Cu\ K_\alpha$ radiation (= 0.1542 nm), θ is the Bragg's diffraction angle and $D$ is the crystallite size. Here, the value of FWHM has been used from Rietveld refinement of corresponding XRD patterns. According to Rietveld method [33], the individual contributions to the broadening of reflections can be expressed as,

$$FWHM^2 = (U + D_{ST}^2)(\tan^2\theta) + V(\tan\theta) + W + \frac{IG}{Cos^2\theta} \quad \text{------------- (2)}$$

Where $U$, $V$ and $W$ are the usual peak shape parameters, $IG$ is a measure of the isotropic size effect, $D_{ST}$ =coefficient related to strain. As shown in the table 1, the crystallite size decreases with the increase of substituent ion concentration. This implies the development of lattice strain inside the lattice due to ionic size mismatch between $La^{3+}$ & $Bi^{3+}$ and $Fe^{3+}$ & $Ti^{4+}$ which leads to local structural disorder and reduces the rate of nucleation as an effect decrease the crystallite size.

All the XRD patterns were analyzed by employing Rietveld method. The refinement procedure is described in previous section (methods). It is worth noting that Rietveld analysis has been carried out for rhombohedral structure of *R3c* space group with atomic positions of Bi/La and Fe/Ti at 6a and O at 18b and orthorhombic structure of *Pbnm* space group with atomic positions of Bi/La at 4c, Fe/Ti at 4a and $O_1$ at 4c and $O_2$ at 8d. Figure 3 shows the typical Rietveld refinements of XRD patterns for BLFT-025 and BLFT-05. The Rietveld refinement of BFO XRD pattern has been carried out considering the rhombohedral phase because it has the characteristic doublet of highest intensity peaks (at $2\theta = 31.8^o$ and $32.1^o$). But this doublet peaks gradually merges for the co-substituted BFO which clearly indicates the presence of modified crystallographic symmetry. Rigorous fitting with the different structural models (*R3c*, *R3c* + *Pbnm, R3m* + *Pbnm, R3c* + *Pbam, R3c* + *P4mm, R3c* + *Pm3m, R3m* + *Pbnm,* etc.) were carried out. The best fit was considered on the basis of goodness of fit parameters such as $\chi^2$, $R_p$, $R_{\omega p}$, $R_{bragg}$ etc. It was observed that BFO (parent compound) crystallizes to rhombohedral symmetry with *R3c* space group. However, XRD patterns of co-substituted ceramics are a result of the superposition of two different crystal symmetries such as *R3c* and *Pbnm*. It indicates that the induced lattice strain due to ionic size mismatch of cations co-substituted BFO leads to the coexistence of rhombohedral (*R3c* space group) and orthorhombic (*Pbnm* space group) crystal symmetries. The percentages of respective contributions of both crystallographic phases have been mentioned in the Table 2.

Fig. 4(a) shows the FE-SEM image for BLFT-015 which shows that the particles have an almost homogeneous distribution of crystallite size. The average crystallite size was found to be ~60 nm for BLFT-015 which is close to the value obtained from XRD patterns (63 nm). The crystallite size decreases with the increase in co-substitution concentration. In addition to it, the substitution of high valence $Ti^{4+}$ suppress the oxygen vacancies which results in decrease of grain size of co-substituted ceramics because it is facilitated by oxygen ions motion. The elemental analysis of ceramics was carried out using the EDS. The typical EDS pattern of BLFT-015 ceramics has been shown in the Fig. 4(b). It reveals the presence of Bi, Fe, La, Ti and O elements in the ceramics and it showed the atomic ratio of bismuth ($Bi^{3+}$) and iron ($Fe^{3+}$) cations as approximately 1:1 and same for lanthanum ($La^{3+}$) and titanium ($Ti^{4+}$) cations. No extra peaks have been traced which indicate that there is no contamination in the ceramics. The TEM image (Fig. 4(c)) for BLFT-015 shows the average crystallite size to be about 60 nm which is consistent with the FE-SEM as well as XRD analysis. Also, SAED pattern of a single crystallite in BLFT-015 (Fig. 4(d)) suggests the particle to be single crystal. The similar kind of SAED patterns were obtained for particles in other region of sample.

Since Micro-Raman spectroscopy is sensitive to the structural phase transitions, it has been carried out to further support the Rietveld analysis of the XRD patterns. In principle, any change in the crystal structure can be studied by examining the variation in frequency, bandwidth and intensity of the Raman peaks. Theoretical group analysis has predicted 18 optical phonon modes: $4A_1+5A_2+9E$ for BFO (space group as *R3c*) at room temperature. The $A_1$ and E modes are both Raman and IR-active modes, whereas the $A_2$ modes are Raman and IR inactive modes [34]. $A_1$ modes are polarized along z-axis whereas E modes in x-y plane. By deconvoluting the fitted curves into individual modes, the peak position of each component, i.e. the natural frequency ($cm^{-1}$) of each Raman active mode, has been obtained for all the Raman spectra. Figure 5(a) and (b) show the typical Raman scattering spectra of BFO with its phonon modes deconvoluted into individual lorentzian components and combined spectra for $Bi_{1-x}La_xFe_{1-x}Ti_xO_3$ (x = 0.000 - 0.250) respectively. All the allowed peaks for BFO have been observed except for the $A_1$ mode below 100 $cm^{-1}$, which cannot be observed due to rejection of the notch filter in the spectrometer. It shows three sharp peaks at around 140, 171 and 220 $cm^{-1}$ could be assigned as $A_1$-1, $A_1$-2 and $A_1$-3 phonon modes respectively. Nine other phonon modes are located in the range of 230 – 700 $cm^{-1}$, the other modes are in the E-symmetry except the $A_1$ mode at 433 $cm^{-1}$ [35]. The Raman peaks below 250 $cm^{-1}$ shift to higher frequency and broaden gradually with increase in the substitution concentration which is related to Bi–O vibrations. The average mass of the A-sites decreases with the increase in substitution concentration because atomic mass of La is 33 % less than that of Bi which leads to shift of $A_1$-1 and $A_1$-2 phonon modes to higher frequency side as the frequency of the phonon mode is inversely proportional to the reduced mass as frequency of the mode is proportional to $(k/M)^{1/2}$, where, k is the force constant and M is the reduced mass. The low frequency modes are attributed to the relative motion of A-site cations against the oxygen octahedrons and hence, these displacements also indicate that the substituent replaces the host cations in BFO lattice. This phenomenon is due to the chemical pressure-induced bond shortening and lattice distortion which arises due to ionic size mismatch between host and substituent cations. The Raman modes for BFO over 200 $cm^{-1}$ are attributed to the stretching and bending modes of the $FeO_6$ octahedra [36]. The evolution of the $E_4$ mode (~

470 cm$^{-1}$) can be interpreted by the changes in the displacement of B-site cations and the octahedral tilts. These results indicate that Ti is being substituted only at Fe sites in the BFO lattice. The above observations suggest the compositional driven structural phase transition from the rhombohedral to orthorhombic symmetry. Similar results about the structural phase transition in bismuth ferrite by applying external high pressure using Mao-Bell diamond anvil cell have been reported by Y. Yang et al [37].

The Fig. 6 shows the FT-IR spectra of the present ceramics for the wave number range of 400-730 cm$^{-1}$. The absorption peaks at ~449 and ~552 cm$^{-1}$ may be attributed to the Fe-O bending and stretching vibrations respectively which are the characteristics of the octahedral $FeO_6$ group in the perovskite compounds [38]. The gradual shift in the frequency of the Fe-O stretching mode indicates the formation of the solid solution. It indicates that substituent ions at the corresponding sites in BFO. The absorption peak around 668 cm$^{-1}$ corresponds to the water vapor. In fact, the absorption peaks corresponding to Fe-O stretching vibration shifts to 566 cm$^{-1}$ emerges with the increase in substituents percentage which indicates the composition-driven structural transitions due to the ionic size mismatch between substituent and host cations. These results are consistent with the XRD as well as Raman spectra analysis.

In order to understand the effect of chemical pressure on the band gap of BFO, the diffuse reflectance spectra of all the ceramics have been recorded. UV-Vis diffuse reflectance spectra were converted into absorption readings according to the Kubelka-Munk (K-M) method [39]. The absorption spectrum of the ceramics transformed from the diffuse reflection spectra using K-M function,

$$F(R) = \frac{(1-R)^2}{2R} \qquad \text{-------------} (4)$$

Where, R is diffuse reflectance. Typical plots of F(R) versus wavelength (λ) have been shown in Fig. 7. By extrapolating F(R) to zero, the direct band gap value have been estimated at 300 K as 2.174, 2.146, 2.141, 2.102, 2.096, 2.078 & 2.072 eV for BFO, BLFT-025, BLFT-05, BLFT-10, BLFT-15, BLFT-20 & BLFT-25 ceramics respectively. The energy band gaps of co-substituted ceramics decreases with the increase in the substitution percentage due to increased internal chemical pressure. The absorption spectra show that all compounds band gaps are in visible light range, suggesting their potential applications as visible-light photocatalysts.

Since BFO has a distorted cubic perovskite structure, there is a point group symmetry breaking from $O_h$ to $C_{3v}$ [40]. There are expected six transitions between 0 and 3 eV by considering $C_{3v}$ local symmetry of $Fe^{3+}$ ions (High spin configuration $t_{2g}^3 e_g^2$ ) in BFO and using the correlation group and subgroup analysis for the symmetry breaking from $O_h$ to $C_{3v}$ [41]. All six transitions have been observed which lie in the range between 1.3 to 3 eV for BFO as shown in the Fig. 8. However, these peaks broaden and vanish gradually with the increase in the substitution concentration and there are only five peaks (of lower relative intensity) in BLFT-025. This indicates the structural transition from the distorted perovskite structure to orthorhombic symmetry due to internal chemical pressure due to size mismatch between substitution and host cations which corresponds to the modification in $FeO_6$ local environment. The analysis of XRD patterns, Raman spectra and FTIR spectra also support well in this context.

The Fig. 9 shows the room temperature magnetization versus magnetic field (M-H) plots of typical ceramics with a maximum applied magnetic field of ± 90 kOe and magnetic parameters have been enlisted in table 3. In the G-type antiferromagnetic structure of BFO, the magnetic moment of $Fe^{3+}$ cations in are ferromagnetically coupled in pseudo cubic (111) planes but antiferromagnetically between adjacent planes and it is surrounded by six $O^{2-}$ ions in the common vertex of two adjacent $FeO_6$ octahedra. Magnetic hysteresis loop for BFO shows the linear magnetic field dependence of magnetization which indicates that it is antiferromagnetic material. However, the simultaneous substitution of La and Ti induces weak ferromagnetism which is evident from the enhancement in remnant magnetization and coercivity for the co-substituted BFO. The unsaturated hysteresis loops and presence of small remnant magnetization indicates the signature of antiferromagnetism with weak ferromagnetism due to the canting of antiferromagnetically ordered spins.

The Fe-O-Fe bond angles have been obtained using Diamond 3.2 software which has been mentioned in Table. The bond angle increases upto 5 % of co-substitution and then decreases. It increases the canting angle of the antiferromagnetically ordered adjacent planes with the suppression of cycloid spin structure [42]. Hence, the evolution of weak ferromagnetism in substituted ceramics may be attributed to the canting of antiferromagnetically ordered spins because of structural distortion as there is no contribution of magnetization from the impurity phases ($Bi_2Fe_4O_9$ and $Bi_{25}FeO_{40}$) because these are paramagnetic at room temperature [43] and also the substituents are non-magnetic. The maximum and remnant magnetization increases with increasing the substitution concentration up to 10% due to suppression of spin cycloid structure. But, the increase in substitution concentration more than 10% breaks the magnetic exchange network in BFO lattice because of nonmagnetic nature of both substituents (La & Ti) and results in the further decrease in both magnetization values. Also, the maximum magnetization at 9 T and remnant magnetization for more than the co-substitution of 10% decreases significantly due to the appearance of collinear antiferromagnetic ordering in the orthorhombic structure because of the significant contribution from the crystallographic phase of *Pbnm* space group (as obtained from the quantitative crystallographic phase contribution by double phase Rietveld analysis).

The magnetic impurities such as $Fe_2O_3$ or $Fe_3O_4$ can affect the magnetic properties of the BFO ceramics. In fact, the presence of $Fe_2O_3$ magnetic impurity has not been detected form XRD (X-Ray diffraction) pattern analysis in the present ceramics. Moreover, the observed magnetic hysteresis loop for BFO which has largest amount of impurity phases among all ceramics, does not show the signature of ferromagnetism. Also, the area of ferromagnetic hysteresis loop and coercivity increases with the increase in substitution percentage upto 10 % and then decreases significantly due to appearance of collinear antiferromagnetic ordering in the orthorhombic structure. These results are also consistent with our crystallographic phase percentage obtained from Rietveld analysis of XRD patterns which shows that the percentage of the orthorhombic crystal symmetry (space group *Pbnm*) is significant for more than 10 % of co-substitution. This result can be attributed to the fact that the weak ferromagnetism in the substituted ceramics might be due to broken cycloid spin structure which leads to canting of the antiferromagnetic spin structure.

Figure 10 shows the room temperature dielectric constant as well as loss versus frequency plots of all ceramics in the frequency range of 300 Hz – 10 MHz. BFO show the large dispersion at low frequency due to significant oxygen vacancies and secondary phases than that for the substituted ceramics. The dielectric constant increases with the increase in substitution percentage and attains maximum for BLFT-10 ceramics, then decreases for the higher substitution due to increase in lattice defects. At the same time, dielectric loss was observed to be decreasing with the increase in co-substitution percentage which can be correlated with the increase in the resistivity because Ti substitution on B-site (i.e. for $Fe^{3+}$) eliminate oxygen vacancies (responsible for low resistivity). Moreover, smaller grain size results in the increase of grain boundaries acting as scattering centre for the moving electrons and hence helps in increasing the resistivity. The decrease of dielectric constant with the increase in frequency can be explained by dipole relaxation phenomenon due to Maxwell–Wagner type of interfacial polarization contribution. The dipoles having large effective masses (e.g., oxygen vacancies, segregated impurities phases) are able to follow at low frequency of the applied field and ceases to respond at higher frequencies. Hence, the dielectric constant decreases rapidly with the frequency in lower frequency range (< $10^4$ Hz) and become frequency independent in higher frequency range (> $10^4$ Hz). The 10 % co-substituted samples shows significant frequency independent region for dielectric constant and dielectric loss whereas BFO shows high frequency dependence. These results indicates the reduced dielectric loss which leads to the decrease in the room temperature bulk conductivity (decrease in leakage current) and thus enhanced resistivity in BLFT-10 ceramics.

## 4. Conclusions

The co-substitution of La and Ti in BFO results in structural transition from rhombohedral symmetry (space group *R3c*) to orthorhombic (space group *Pbnm*) as indicated by the XRD as well as Raman spectra analysis. It enhances the resultant magnetic moment due to the broken cycloid spin structure caused by the distortion in the crystal lattice. The crystallographic phase percentage have been quantified by double phase Rietveld analysis of all XRD patterns which shows that the orthorhombic crystal symmetry (space group *Pbnm*) is significant beyond 10% of co-substitution which results in the reduction of the $M_S$, $M_r$ and $H_C$ in this case due to appearance of collinear antiferromagnetic ordering in the orthorhombic crystal structure. The improved dielectric properties with very low value of dielectric loss (0.03) indicate the lowest oxygen vacancies for 10% of co-substitution. The band gap energy of co-substituted ceramics decreases with the increase in the substitution percentage due to increased lattice strain. UV-Visible absorption spectra analysis also supports the modification in local $FeO_6$ environment and structural transition. The enhanced magnetic and dielectric properties may be interesting for multiferroic devices applications.

## Acknowledgment

The authors gratefully acknowledge Dr. Dhanvir Singh Rana and his research group at IISER Bhopal for extending the MPMS facility. We are also grateful to Dr. N. P. Lalla at UGC-DAE-CSR, Indore for his kind help in TEM measurements.

**References**

[1] Eerenstein W, Mathur ND and Scott JF (2006) Multiferroic and magnetoelectric materials, Nature 442: 759-765.

[2] Catalan G and Scott JF (2009) Physics and Applications of Bismuth Ferrite, Adv Mater 21: 2463-2485.

[3] Cheong SW and Mostovoy M (2007) Multiferroics: a magnetic twist for ferroelectricity, Nature Mater 6: 13-20.

[4] Ramesh R and Spaldin NA (2007) Multiferroics: progress and prospects in thin films, Nature Mater 6: 21-29.

[5] Singh A, Pandey V, Kotnala RK and Pandey D (2008) Direct Evidence for Multiferroic Magnetoelectric Coupling in $0.9BiFeO_3$–$0.1BaTiO_3$, Phys Rev Lett 101: 247602-247606.

[6] Ederer C and Spaldin NA (2005) Weak ferromagnetism and magnetoelectric coupling in bismuth ferrite, Phys Rev B 71: 060401-060405.

[7] Dzyaloshinsky I (1958) A thermodynamic theory of “weak” ferromagnetism of antiferromagnetics, J Phys Chem Solids 4: 241-255.

[8] Moriya T (1960) Anisotropic Superexchange Interaction and Weak Ferromagnetism, Phys Rev 120: 91-98.

[9] Sosnowska I, Peterlin-Neumaier T and Steichele E (1982) Spiral magnetic ordering in bismuth ferrite, J Phys C 15: 4835 - 4846.

[10] Kadomtseva AM, Popov Yu F, Pyatakov AP, Vorob’ev GP, Zvezdin AK and Viehland D (2006) Phase Transit: A Multinational Journal, Phase Transit: A Multinational Journal, 79 (12): 1019-42.

[11] Bras Le G, Colson D, Forget A, Genand-Riondet N, Tourbot R and Bonville P (2009) Magnetization and magnetoelectric effect in $Bi_{1-x}La_xFeO_3$ ($0 \leq x \leq 0.15$), Phys Rev B 80: 134417- 6.

[12] Wang N, Cheng J, Pyatakov A, Zvezdin AK, Li JF, Cross LE and Viehland D (2005) Multiferroic properties of modified BiFeO3-PbTiO3-based ceramics: Random-field induced release of latent magnetization and polarization, Phys Rev B 72: 104434-9.

[13] Zalesskii AV, Frolov AA, Khimich TA and Bush AA (2003) Composition-induced transition of spin-modulated structure into a uniform antiferromagnetic state in a $Bi_{1-x}La_xFeO_3$ system studied using 57Fe NMR, Phys Solid State 45: 141-145.

[14] Palkar VR, Kundaliya DC, Malik SK and Bhattacharya S (2004) Magnetoelectricity at room temperature in the $Bi_{0.9-x}Tb_xLa_{0.1}FeO_3$ system, Phys Rev B 69: 212102-212105.

[15] Xu Q, Zai H, Wu D, Tanga YK and Xu MX (2009) The magnetic properties of BiFeO3 and $Bi(Fe_{0.95}Zn_{0.05})O_3$, J Alloys Compd 485: 13-16.

[16] Wei J, Xue D, Wu C and Li Z (2008) Enhanced ferromagnetic properties of multiferroic $Bi_{1-x}Sr_xMn_{0.2}Fe_{0.8}O_3$ synthesized by sol–gel process, J Alloys Compd 453: 20-23.

[17] Cheng ZX, Li AH, Wang XL, Dou SX, Ozawa K, Kimura H, Zhang SJ and Shrout TR (2008) Structure, ferroelectric properties, and magnetic properties of the La-doped bismuth ferrite, J Appl Phys 103: 07E507-3.

[18] Abe K, Sakai N, Takahashi J, Itoh H, Adachi N, Ota T (2010) Leakage Current Properties of Cation-Substituted $BiFeO_3$ Ceramics, Jpn. J Appl Phys, 49: 09MB01-09MB01-6.

[19] Qi X, Dho J, Tomov R, Blamire MG and MacManus-Driscoll JL (2005) Greatly reduced leakage current and conduction mechanism in aliovalent-ion-doped BiFeO3, Appl Phys Lett, 86: 062903-062903-3.

[20] Srinivas A, Kin DW, Hong KS and Suryanarayana SV (2003) Observation of ferroelectromagnetic nature in rare-earth-substituted bismuth iron titanate, Appl Phys Lett, 83: 2217-2219.

[21] Cheng ZX, Wang XL, Du Y and Dou SX (2010) A way to enhance the magnetic moment of multiferroic bismuth ferrite, J Phys D: Appl Phys. 43: 242001-242006.

[22] Ghosh S, Dasgupta S, Sen A and Maiti HS (2005) Low-Temperature Synthesis of Nanosized Bismuth Ferrite by Soft Chemical Route, J Am Ceram Soc, 88: 1349-1352.

[23] Kumar MM, Palkar VR, Srinivas K and Suryanarayana SV (2000) Ferroelectricity in a pure $BiFeO_3$ ceramic, Appl Phys Lett 76: 2764-2766.

[24] Pradhan AK, Zhang K, Hunter D, Dadson JB, Loutts GB, Bhattacharya P, Katiyar R, Zhang J, Sellmyer DJ, Roy UN, Cui Y and Burger A (2005) Magnetic and electrical properties of single-phase multiferroic $BiFeO_3$, J Appl Phys 97: 093903-1-093903-4.

[25] Cheng ZX, Li AH, Wang XL, Dou SX, Ozawa K, Kimura H, Zhang SJ and Shrout TR (2008) Structure, ferroelectric properties, and magnetic properties of the La-doped bismuth ferrite, J Appl Phys 103(7): 07E507-07E507-3.

[26] Kumar M and Yadav KL (2006) Study of room temperature magnetoelectric coupling in Ti substituted bismuth ferrite system, J Appl Phys 100: 074111-074111-4.

[27] Lee C-C and Wu J-M (2007) Studies on Leakage Mechanisms and Electrical Properties of Doped $BiFeO_3$, *Solid-State Lett*, 10(8): G58-G61

[28] J Wu and J Wang (2010) Elimination of domain backswitching in $BiFe_{0.95}Mn_{0.05}O_3$ thin films by lowering the layer thickness, J Am Ceram Soc 93: 2795–2803.

[29] P Kumar and M Kar (2014) Effect of structural transition on magnetic and optical properties of Ca and Ti co-substituted $BiFeO_3$ ceramics, J Alloys Compd 584: 566-572.

[30] J Rodriguez-Carvajal Laboratory, FULLPROF (2010) a Rietveld and pattern matching and analysis programs version, Laboratoire Leon Brillouin, CEA-CNRS, France.

[31] Palai R, Katiyar RS, Schmid H, Tissot P, Clark SJ, Robertson J, Redfern SAT, Catalan G and Scott JF (2008) β phase and γ-β metal-insulator transition in multiferroic $BiFeO_3$, Phys Rev B 77: 014110-014110-11.

[32] Cullity BD (1978) Elements of X-ray diffraction 2nd ed. Addison-Wesley series

[33] Young RY (1996) The Rietveld Method 3rd ed. Oxford University Press

[34] Kothari D, Reddy VR, Sathe VG, Gupta A, Banerjee A and Awasthi AM (2008) Raman scattering study of polycrystalline magnetoelectric $BiFeO_3$, J Magn Magn Mater 320: 548-552.

[35] Yuan GL, Or SW and Chan HLW (2007) Reduced ferroelectric coercivity in multiferroic $Bi_{0.825}Nd_{0.175}FeO_3$ thin film J Appl Phys 101: 024106-024106-4.

[36] Lin JW, Tite T, Tang YH, Lue CS, Chang YM and Lin JG (2012) Electron spin resonance probed suppressing of the cycloidal spin structure in doped bismuth ferrites, J Appl Phys 111: 07D910-07D910-3.

[37] Yang Y, Bai LG, Zhu K, Liu YL, Jiang S, Liu J, Chen J and Xing XR (2009) High pressure Raman investigations of multiferroic $BiFeO_3$, J Phys.: conds Matt 21: 385901-1-385901-5.

[38] Mishra RK, Pradhan DK, Choudhary RNP and Banerjee A (2008) Effect of yttrium on improvement of dielectric properties and magnetic switching behavior in $BiFeO_3$, J Phys: Condens Matter 20: 045218-045223.

[39] Kubelka P and Munk FZ (1931) Ein Beitrag zur Optik der Farbanstriche, Tech. Phys. 12: 593-601.

[40] Ramirez MO, Kumar A, Denev SA, Podraza NJ, Xu XS, Rai RC, Chu YH, Seide J, Martin LW, Yang SY, Saiz E, Ihlefeld JF, Lee S, Klug J, Cheong SW, Bedzyk MJ, Auciello O, Schlom DG, Ramesh R, Orenstein J, Musfeldt JL and Gopalan V (2009) Magnon sidebands and spin-charge coupling in bismuth ferrite probed by nonlinear optical spectroscopy, Phys. Rev. B 79: 224106-224114.

[41] Ramachanran B, Dixit A, Naik R, Lawes G and Rao MSR (2010) Charge transfer and electronic transitions in polycrystalline $BiFeO_3$, Phys Rev B 82 012102-012105.

[42] Yang CH, Kan D, Takeeuchi I, Nagarajan V and Seidel J (2012) Doping $BiFeO_3$: approaches and enhanced functionality, Phys Chem Chem Phys 14: 15953-15962.

[43] Shamir N, Gurewitz E and Shaked H (1978) The magnetic structure of $Bi_2Fe_4O_9$ - analysis of neutron diffraction measurements, Acta Cryst A34: 662-666.

**Table 1** Structural parameters of $Bi_{1-x}La_xFe_{1-x}Ti_xO_3$ (for x = 0.000 - 0.250) ceramics annealed at 700 $^oC$. Where, x = substitution value, D = crystallites size, $\chi 2$ = Chi-square (goodness of fit).

| X | D | Fe-O-Fe Angle (R3c) | V (Å$^3$) (R3c) | $\chi 2$ |
|---|---|---|---|---|
| 0.000 | 82.49 | 154.487 | 372.470 | 1.39 |
| 0.015 | 60.82 | 154.570 | 372.421 | 1.37 |
| 0.025 | 47.55 | 154.852 | 372.294 | 1.25 |
| 0.050 | 44.51 | 154.997 | 372.161 | 1.31 |
| 0.075 | 43.14 | 155.413 | 372.011 | 1.24 |
| 0.100 | 41.18 | 158.482 | 371.730 | 1.27 |
| 0.150 | 38.62 | 156.311 | 371.252 | 1.32 |
| 0.200 | 27.48 | 148.310 | 370.110 | 1.34 |
| 0.250 | 25.32 | 142.290 | 368.504 | 1.30 |

**Table 2** Crystallographic phase contribution of all ceramics obtained by the Rietveld refinement of XRD patterns for $Bi_{1-x}La_xFe_{1-x}Ti_xO_3$ for x = 0.000 - 0.250.

| Space Group | *R3c* | *Pbnm* |
|---|---|---|
| | **Crystallographic Phase contribution of all ceramics** | |
| X = 0.000 | 100 % | 0 % |
| X = 0.015 | 97.59 % | 2.41 % |
| X = 0.025 | 91.59 % | 8.41 % |
| X = 0.050 | 88.79 % | 11.21 % |
| X = 0.075 | 81.29 % | 18.71% |
| X = 0.100 | 70.37 % | 29.63% |
| X = 0.150 | 53.46 % | 46.54 % |
| X = 0.200 | 48.28 % | 51.72 % |
| X = 0.250 | 30.24 % | 69.76 % |

**Table 3** Magnetic parameters of $Bi_{1-x}La_xFe_{1-x}Ti_xO_3$ (for x = 0.000 - 0.250) ceramics annealed at 700 $^oC$. Where, $M_S$ = magnetization at maximum applied field, $H_C$ = Coercive field, emu/g = emu/gram at applied magnetic field of 90 kOe.

| $Bi_{1-x}La_xFe_{1-x}Ti_xO_3$ | $M_S$ at 90 kOe (emu/g) | $M_r$ (emu/g) | $H_C$ (kOe) |
|---|---|---|---|
| X = 0.000 | 0.6532 | 0.0086 | 0.1701 |
| X = 0.015 | 0.7187 | 0.0139 | 1.6706 |
| X = 0.050 | 0.9105 | 0.1460 | 11.2217 |
| X = 0.100 | 0.9448 | 0.1920 | 12.9889 |
| X = 0.150 | 0.9080 | 0.1788 | 12.9782 |
| X = 0.200 | 0.9078 | 0.1778 | 12.9703 |
| X = 0.250 | 0.8482 | 0.1174 | 8.9532 |

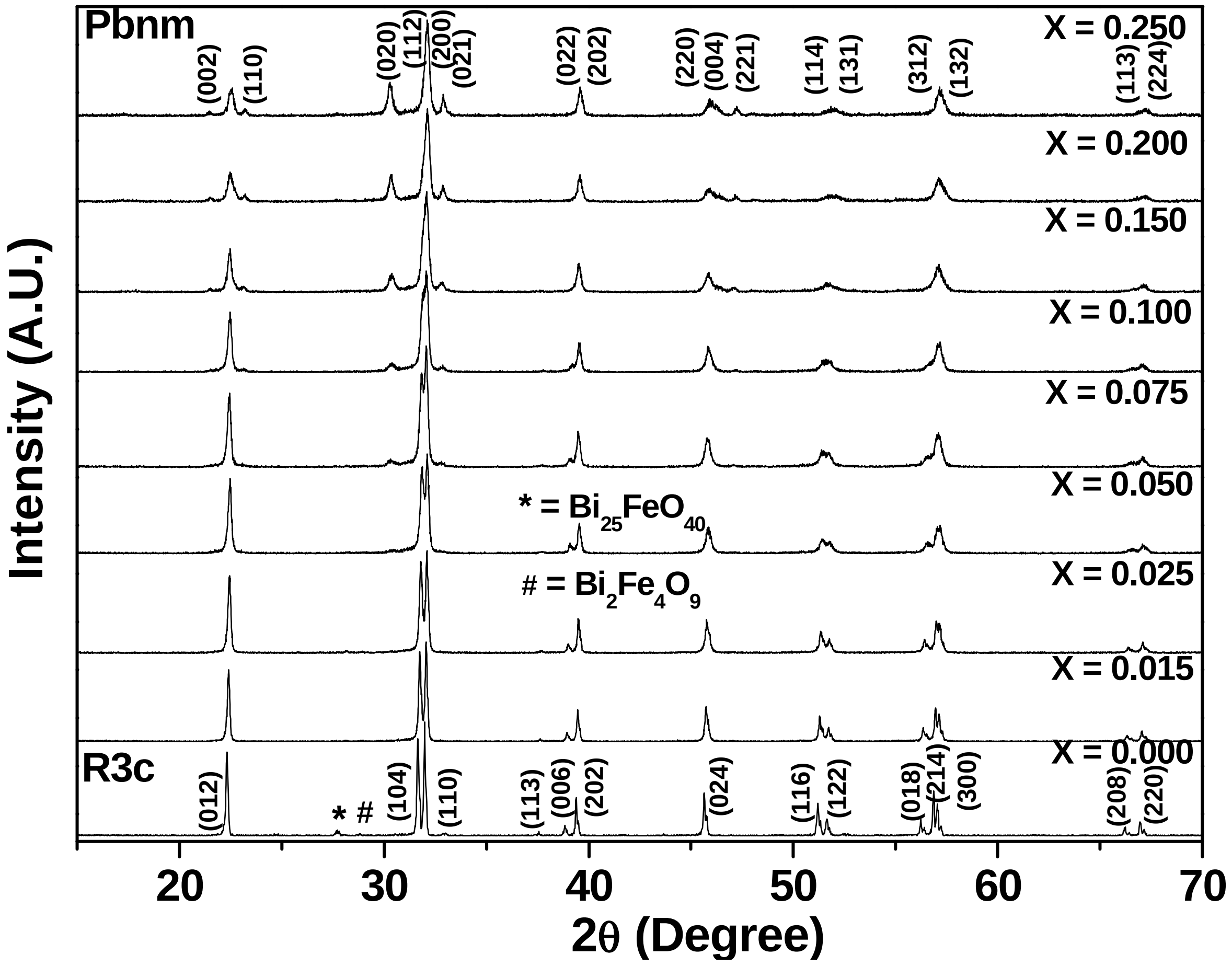

**Fig. 1.** XRD patterns of $Bi_{1-x}La_xFe_{1-x}Ti_xO_3$ ceramics for x = 0.000 - 0.250. The ceramics with x = 0.250 have been indexed according to *Pbnm* space group for clarity but it has contribution from both crystallographic phases as mentioned in Table 2.

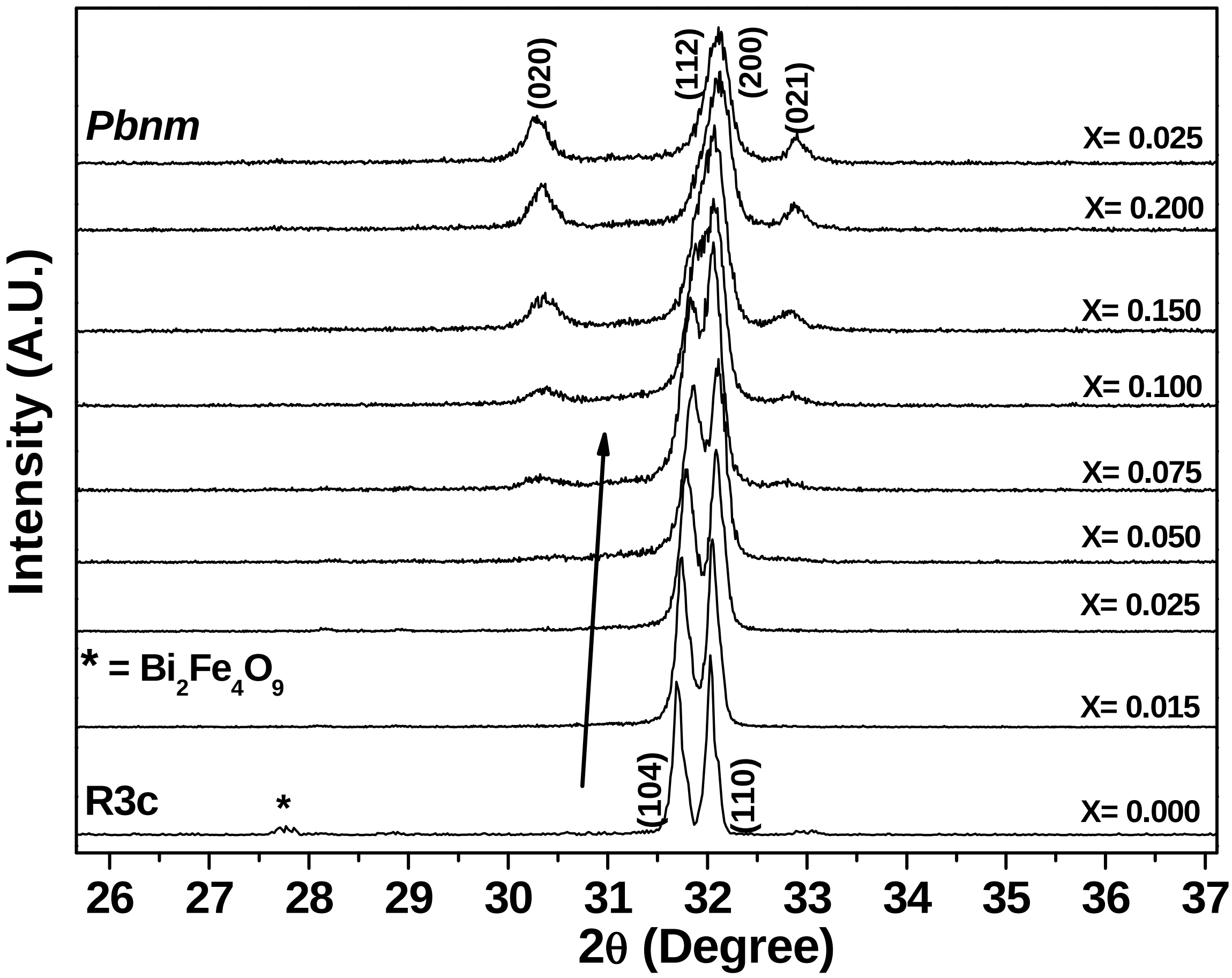

**Fig. 2.** Shift in the XRD patterns of $Bi_{1-x}La_xFe_{1-x}Ti_xO_3$ ceramics for x = 0.000 – 0.250.

(a)

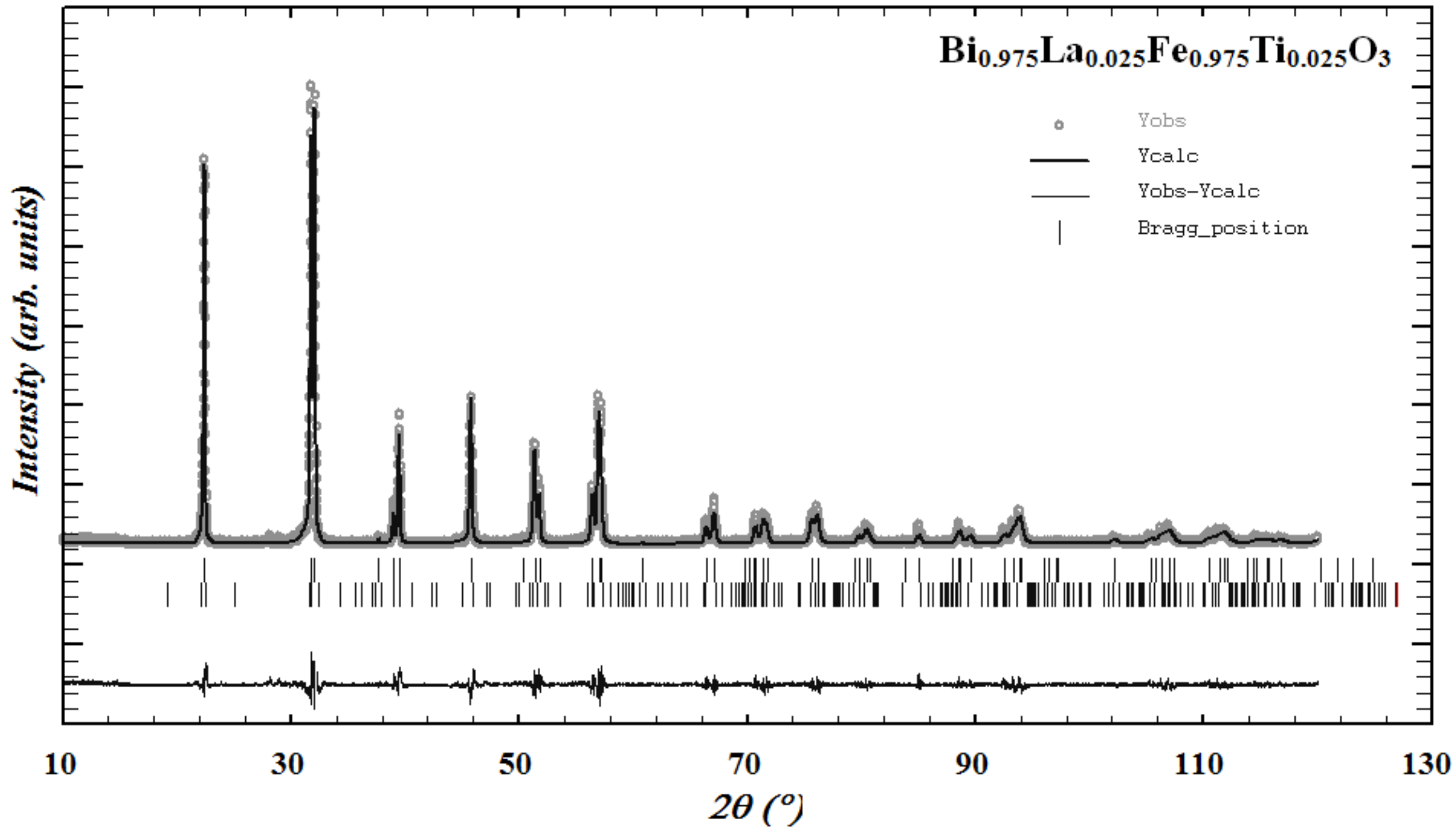

(b)

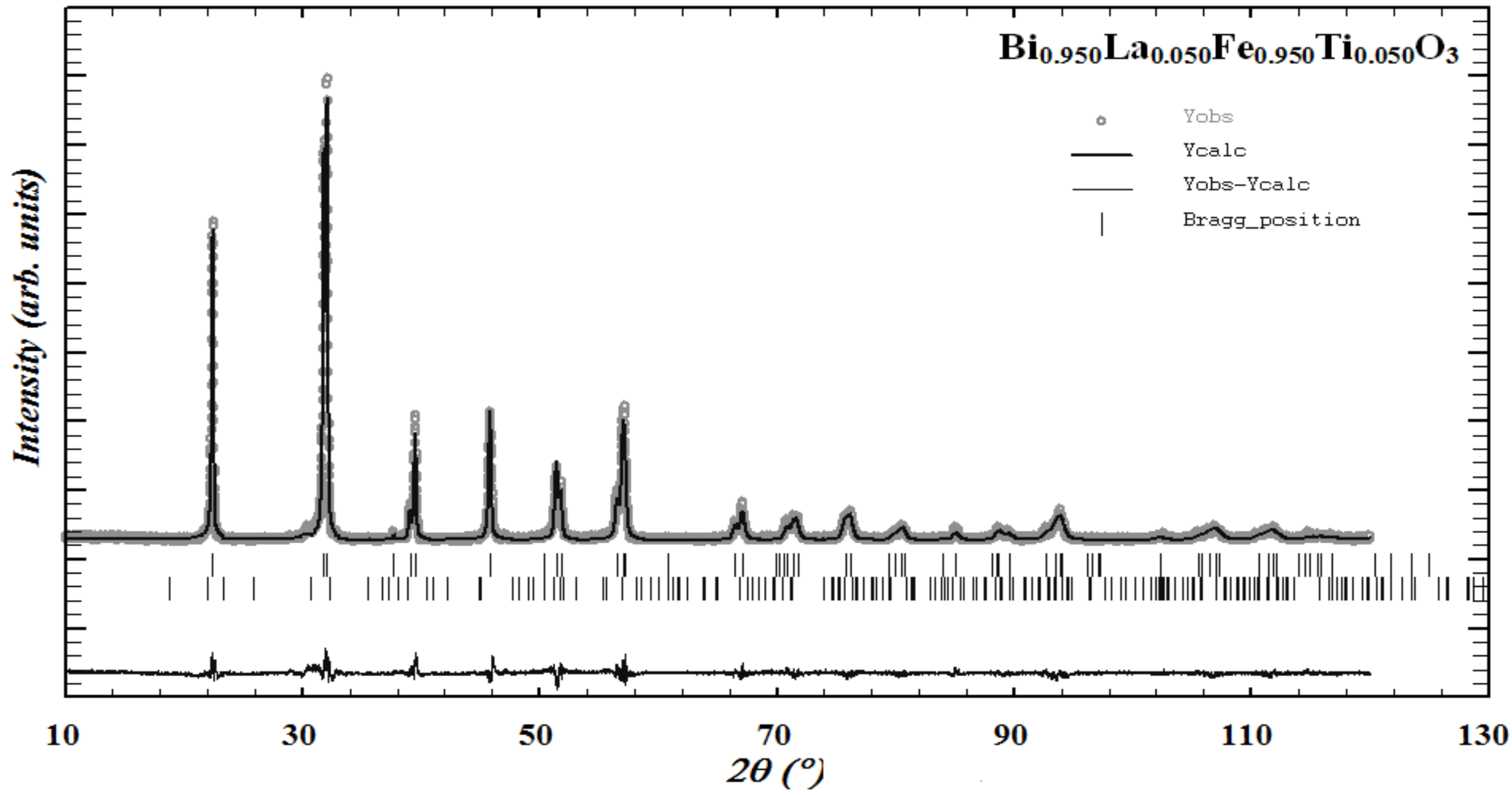

**Fig. 3.** Rietveld refined XRD patterns of $Bi_{1-x}La_xFe_{1-x}Ti_xO_3$ ceramics with (a) x = 0.025 and (b) x = 0.050.

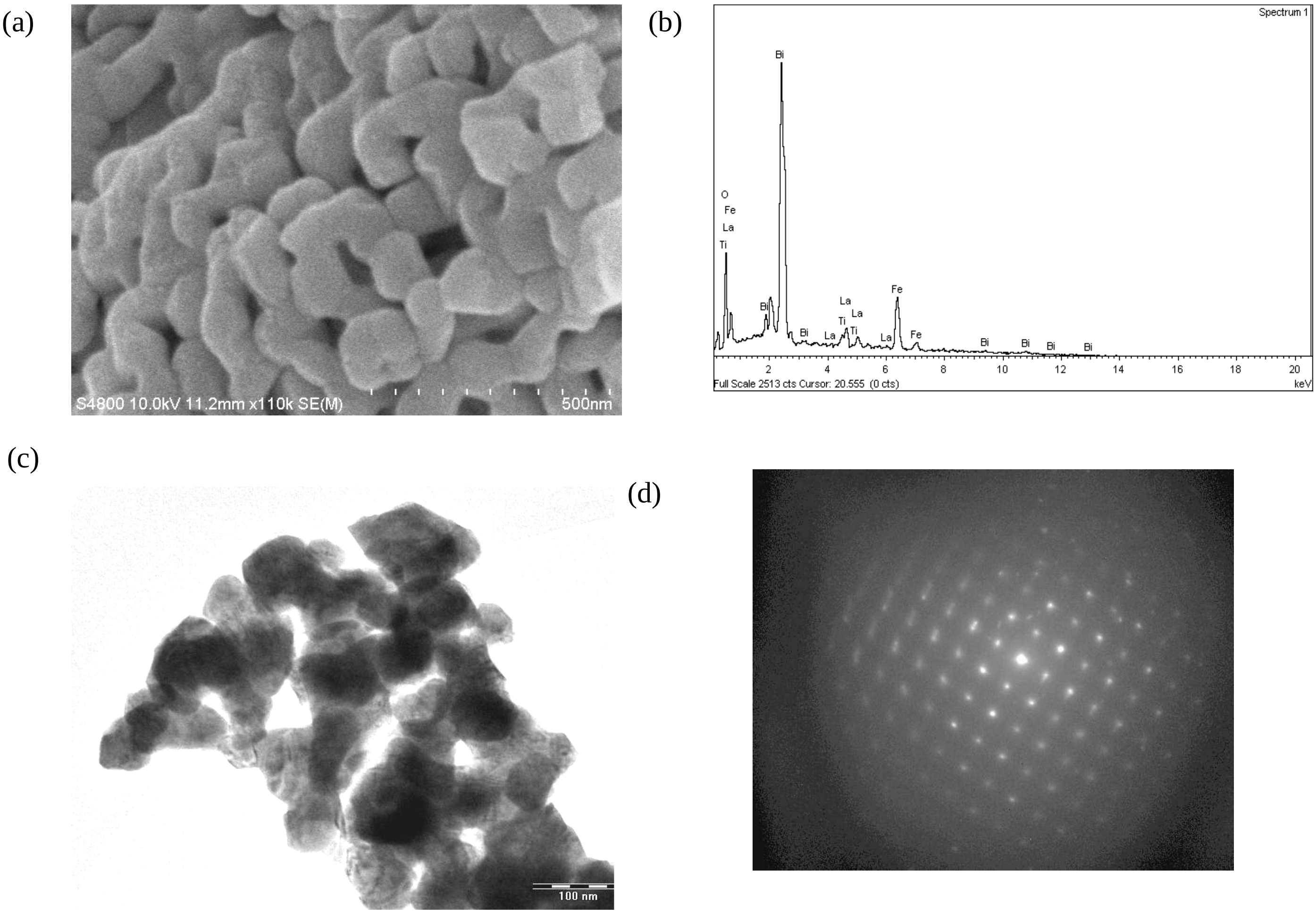

**Fig. 4**. (a) FE-SEM image (b) TEM image and (c) EDS pattern (d) SAED pattern of BLFT-015 ceramics.

(a)
(b)

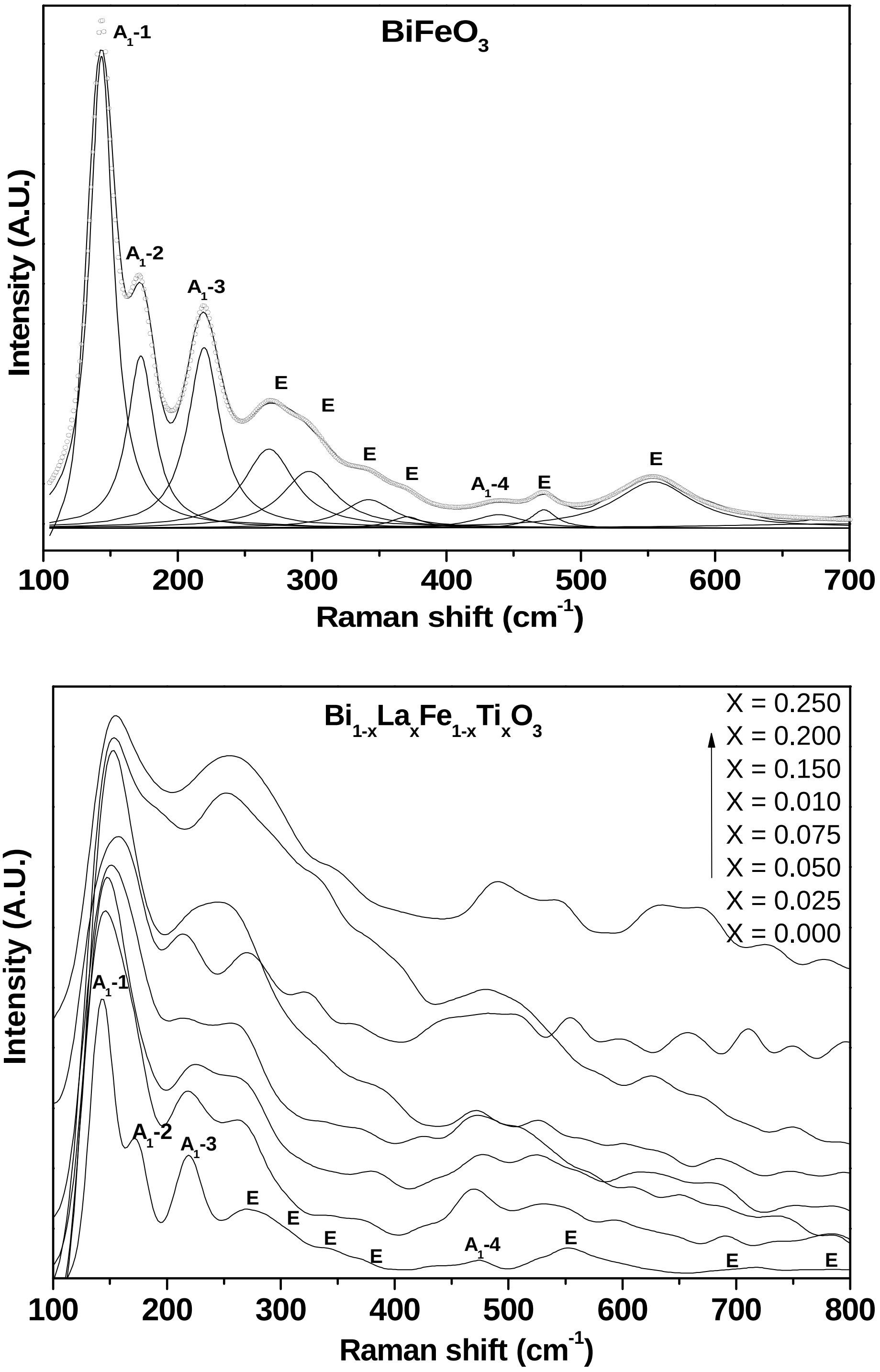

**Fig. 5.** Typical Raman scattering spectra of BFO with its phonon modes deconvoluted into individual lorentzian components and combined spectra of $Bi_{1-x}La_xFe_{1-x}Ti_xO_3$ ceramics for x = 0.000 - 0.250.

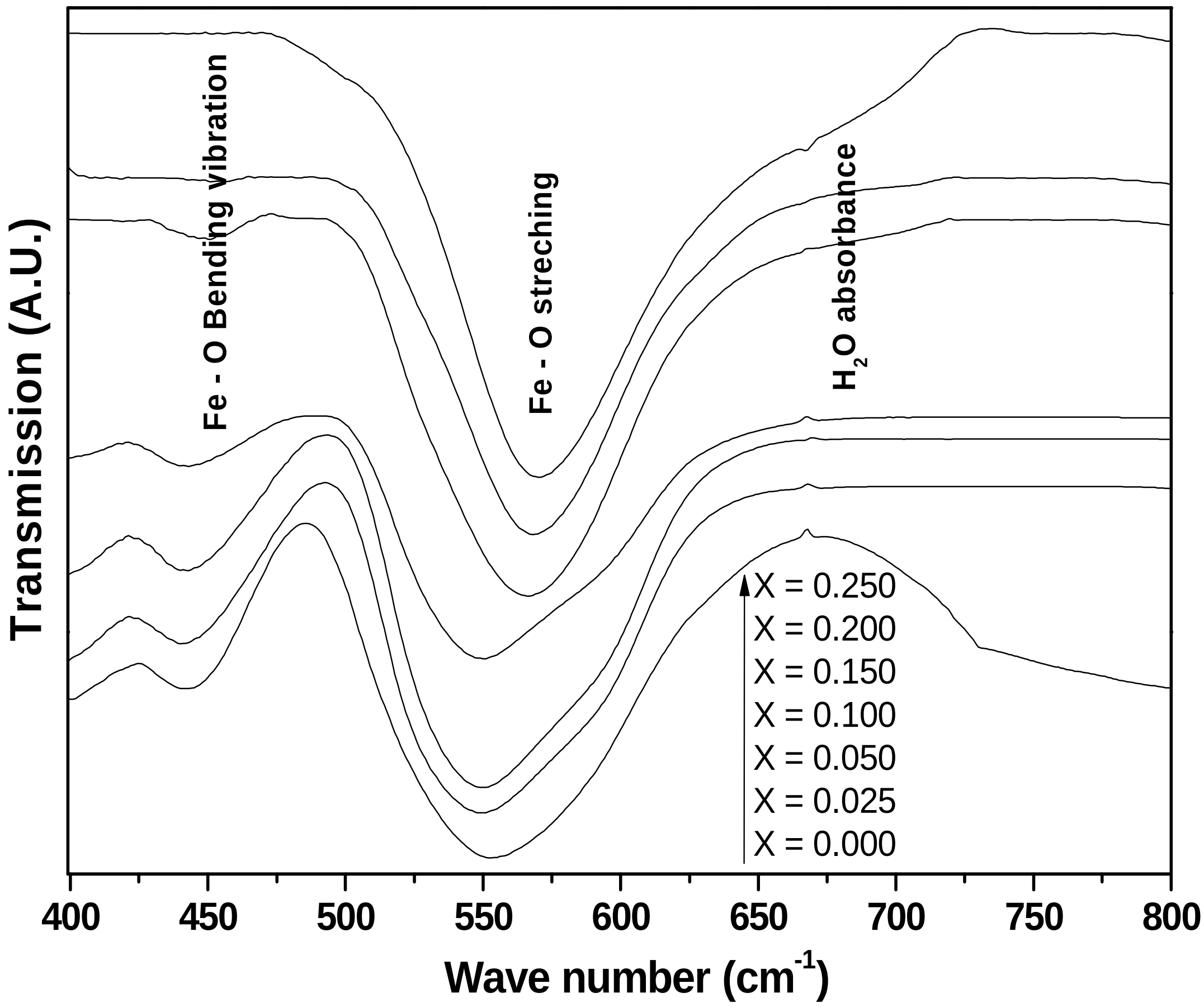

**Fig. 6.** FT-IR spectra of $Bi_{1-x}La_xFe_{1-x}Ti_xO_3$ ceramics for x = 0.000 – 0.250.

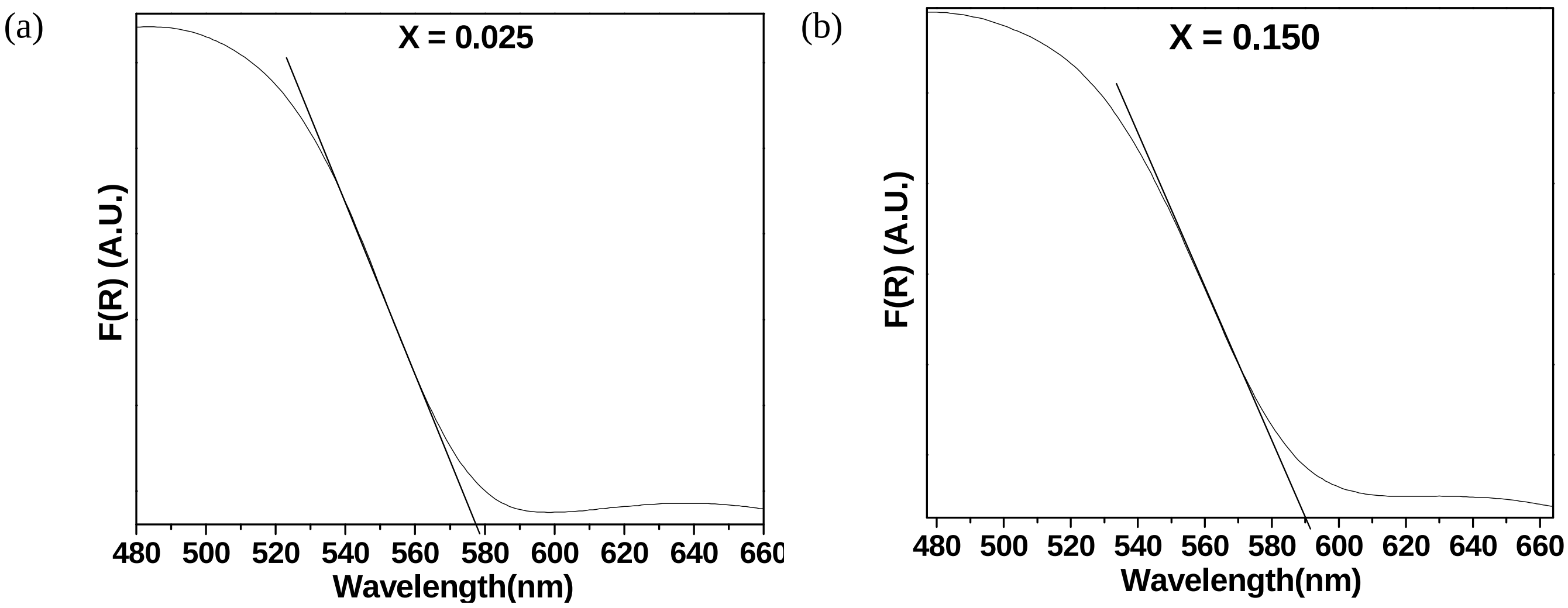

**Fig. 7**. UV-Visible absorption spectra and linear extrapolation of $Bi_{1-x}La_xFe_{1-x}Ti_xO_3$ ceramics with (a) x = 0.025 & (b) x = 0.150, where F(R) = $(1-R)^2/2R$ and R is diffuse reflectance.

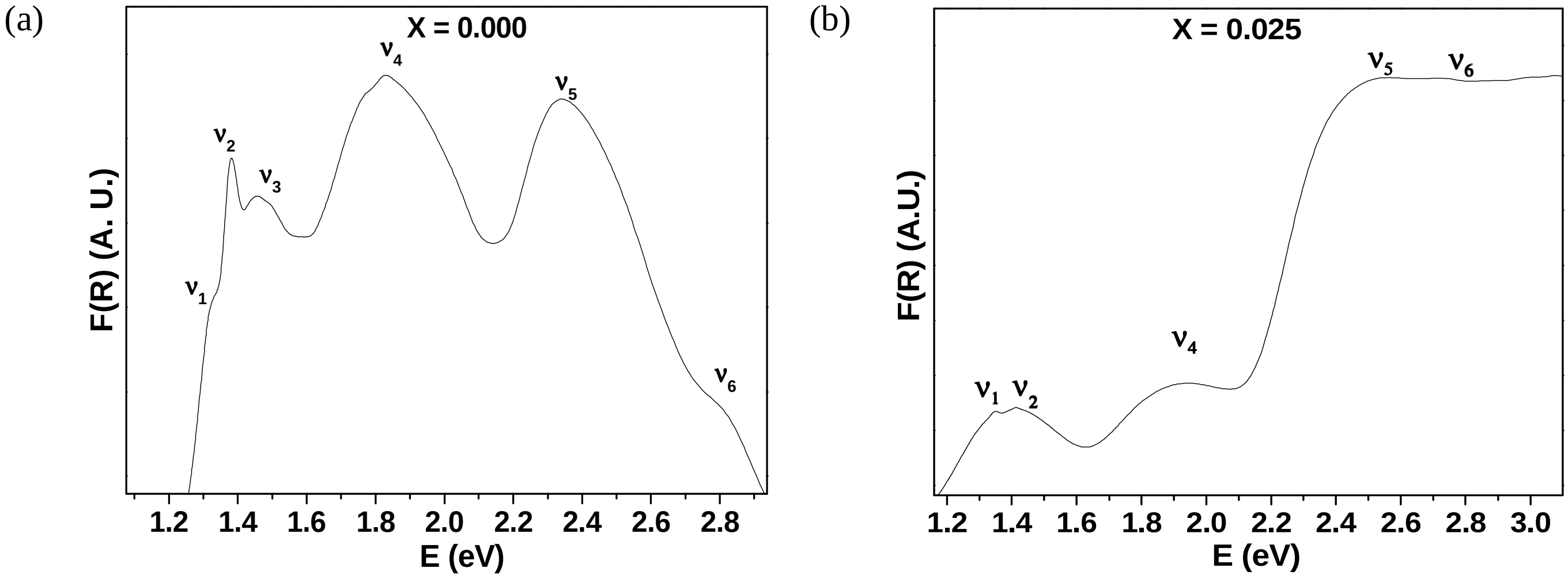

**Fig. 8**. UV-Visible absorption spectra of $Bi_{1-x}La_xFe_{1-x}Ti_xO_3$ ceramics with (a) x = 0.000 & (b) x = 0.025 in the energy range of 1 – 3.2 eV.

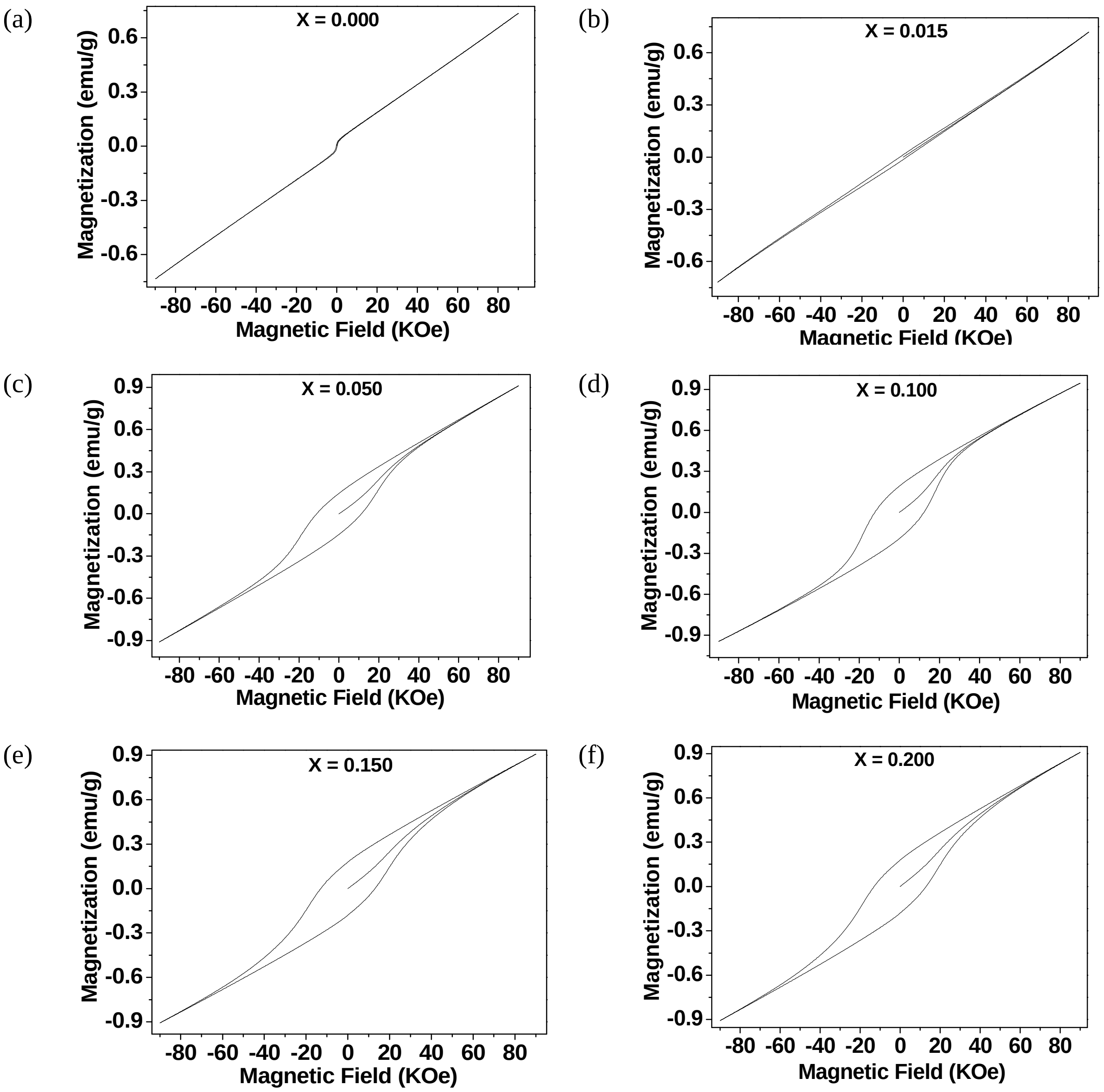

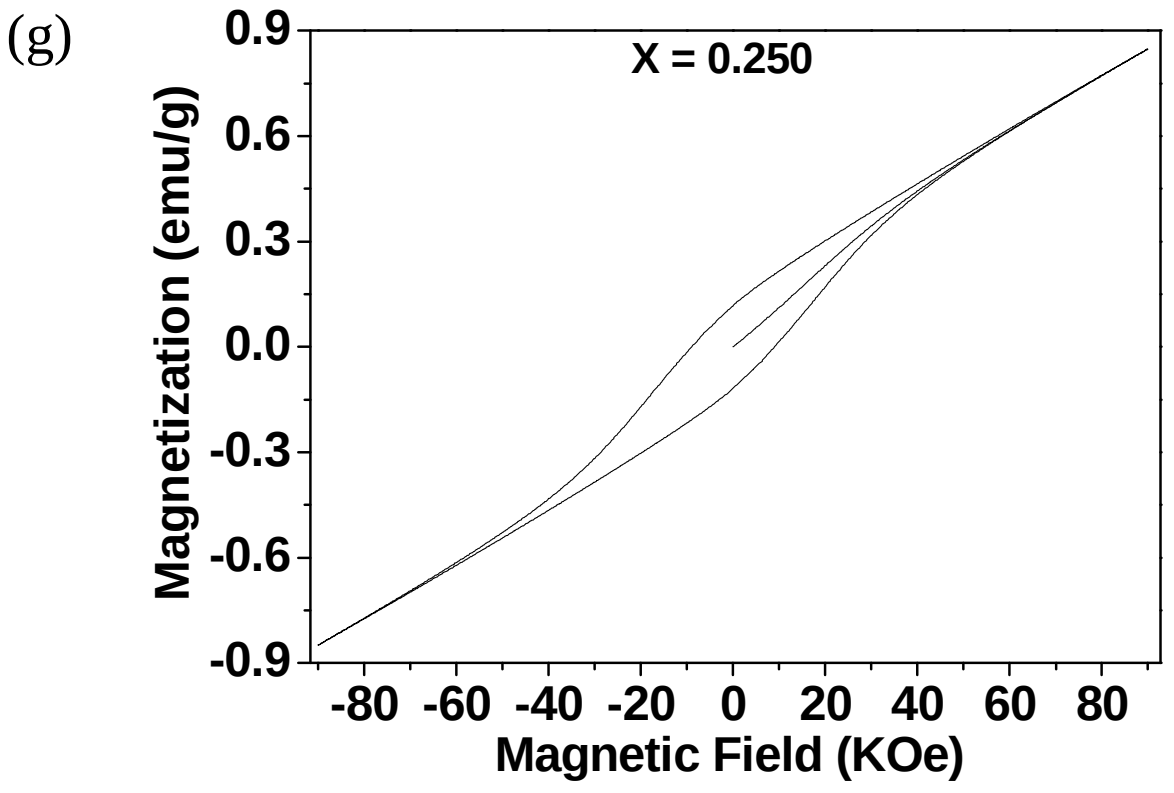

**Fig. 9**. M-H loops of $Bi_{1-x}La_xFe_{1-x}Ti_xO_3$ ceramics with (a) x = 0.000, (b) x = 0.015, (c) x = 0.050, (d) x = 0.100, (e) x = 0.150, (f) x = 0.200, (g) x = 0.250.

(a)

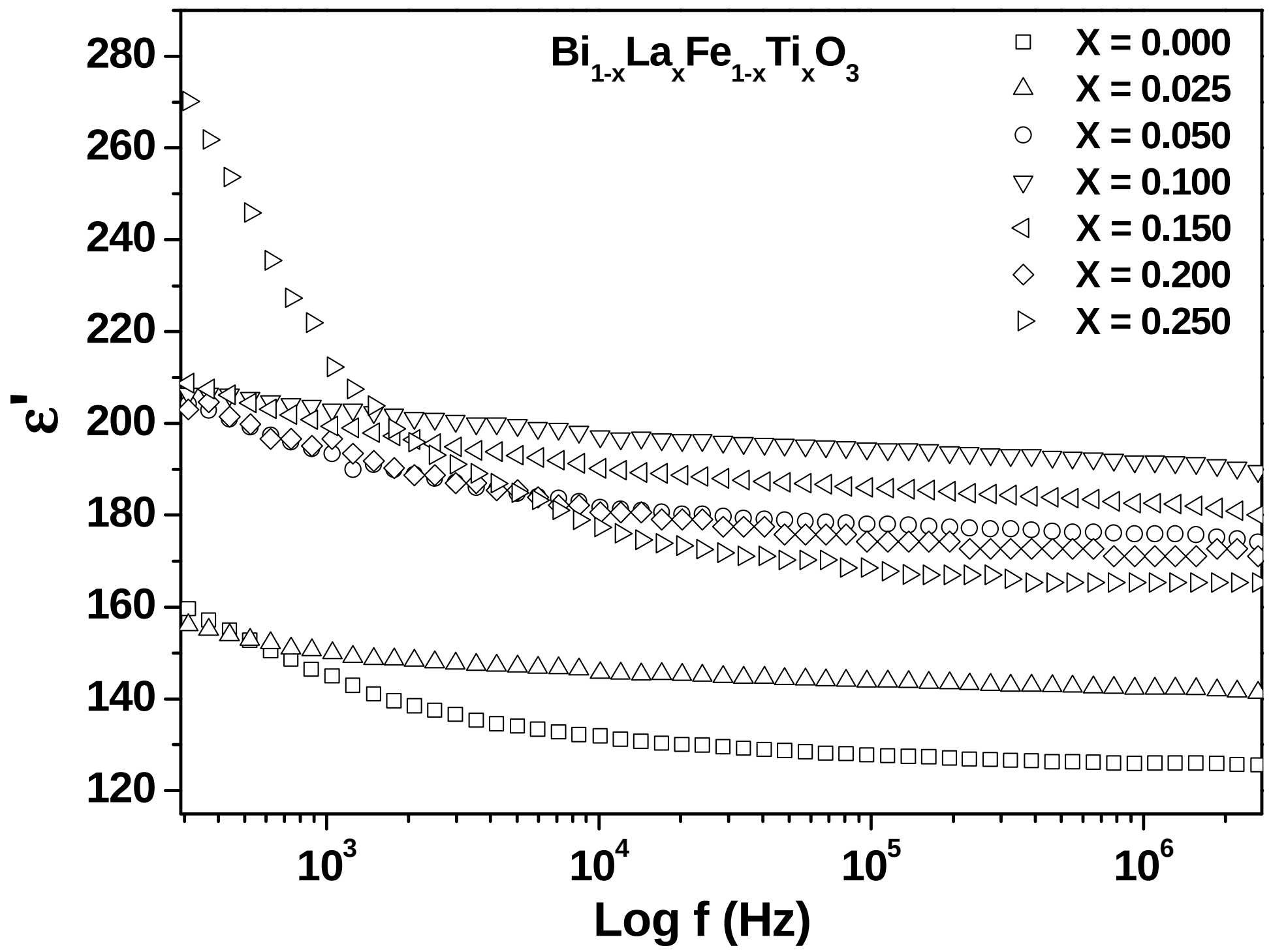

(b)

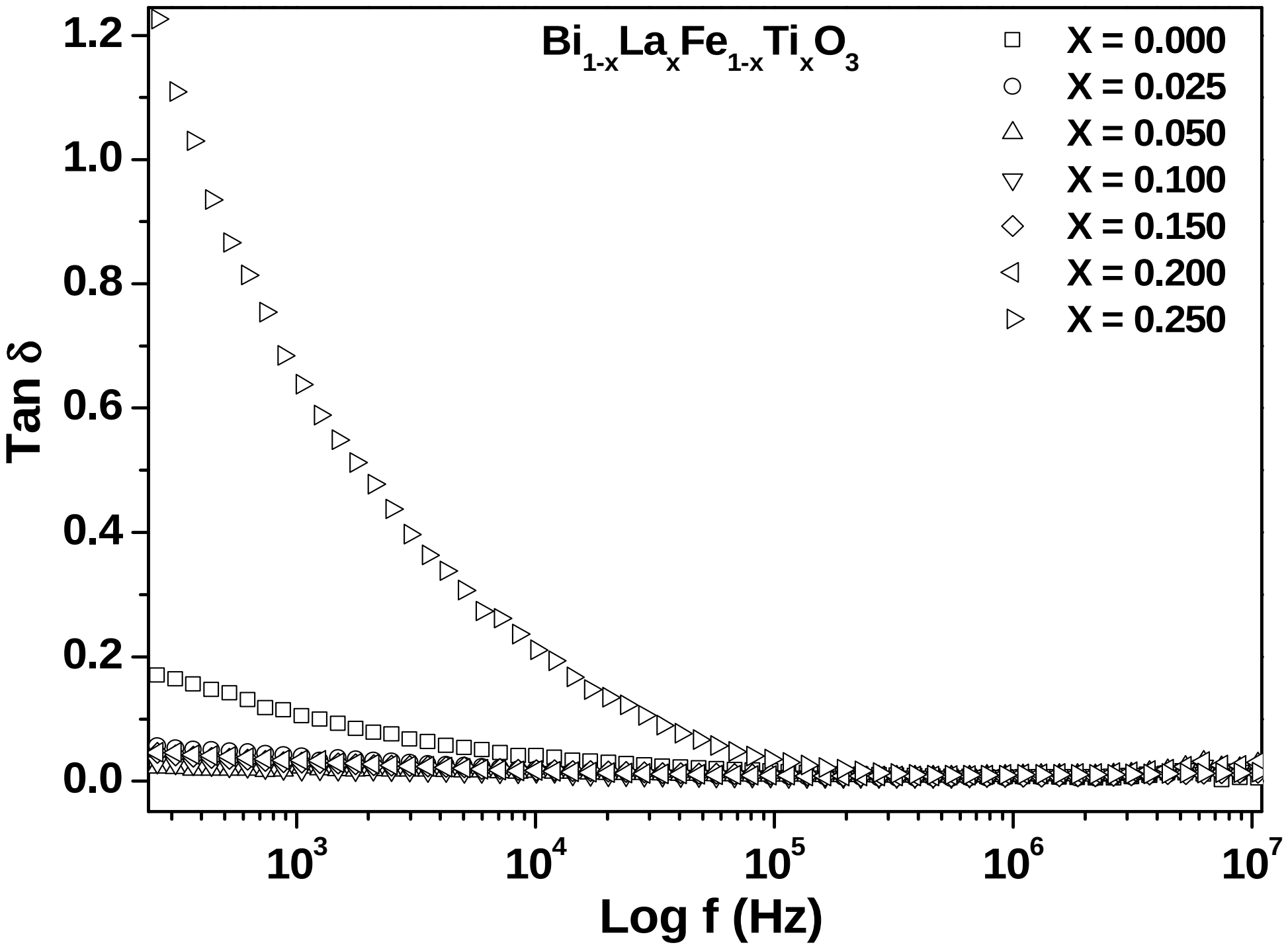

**Fig. 10**. (a) Room temperature dielectric constant and (b) dielectric loss versus frequency (plotted at log scale) plots of $Bi_{1-x}La_xFe_{1-x}Ti_xO_3$ ceramics for x = 0.000 - 0.250.